\documentclass[a4paper,10pt,5p,twocolumn]{elsarticle}
\pdfoutput=1
\usepackage[colorlinks]{hyperref}  
\usepackage{orcidlink}  
\usepackage{siunitx-v2}

\newcommand{\customfootnotetext}[2]{{  
  \renewcommand{\thefootnote}{#1}  
  \footnotetext[0]{#2}}}  
\journal{NIM-A}

\begin{document}

\begin{frontmatter}
\title{Design and Commissioning of Readout Electronics for a $K_L^0$ and $\mu$ Detector at the Belle~II Experiment}
\author{C.~Ketter\orcidlink{0000-0002-5161-9722}}
\author{M.~Andrew\orcidlink{0000-0003-2977-5689}}
\author{T.~Aushev\orcidlink{0000-0002-6347-7055}}
\author{N.K.~Baghel\orcidlink{0009-0008-7806-4422}}
\author{Sw.~Banerjee\orcidlink{0000-0001-8852-2409}}
\author{E.~Becker\orcidlink{0009-0006-4603-3328}}
\author{M.~Beretta\orcidlink{0000-0002-7026-8171}}
\author{E.~Bernieri\orcidlink{0000-0002-4787-2047}}
\author{D.~Biswas\orcidlink{0000-0002-7543-3471}}
\author{D.~Bodrov\orcidlink{0000-0001-5279-4787}}
\author{P.~Branchini\orcidlink{0000-0002-2270-9673}}
\author{A.~Budano\orcidlink{0000-0002-0856-1131}}
\author{C.~Chen\orcidlink{0000-0003-1589-9955}}
\author{Y.~T.~Chen\orcidlink{0000-0003-2639-2850}}
\author{K.~Chilikin\orcidlink{0000-0001-7620-2053}}
\author{S.~Choudhury\orcidlink{0000-0001-9841-0216}}
\author{J.~Cochran\orcidlink{0000-0002-1492-914X}}
\author{G.~De~Pietro\orcidlink{0000-0001-8442-107X}}
\author{R.~de~Sangro\orcidlink{0000-0002-3808-5455}}
\author{G.~Finocchiaro\orcidlink{0000-0002-3936-2151}}
\author{V.~Gaur\orcidlink{0000-0002-8880-6134}}
\author{E.~Graziani\orcidlink{0000-0001-8602-5652}}
\author{Y.~Guan\orcidlink{0000-0002-5541-2278}}
\author{W.~W.~Jacobs\orcidlink{0000-0002-9996-6336}}
\author{S.~Kang\orcidlink{0000-0002-5320-7043}}
\author{T.~D.~Kimmel\orcidlink{0000-0002-9743-8249}}
\author{H.~Kindo\orcidlink{0000-0002-6756-3591}}
\author{B.~Kirby\orcidlink{0000-0001-9432-8880}}
\author{B.~Kunkler$^\ast$\orcidlink{members/001G000001FO17UIAT}}
\author{T.~Lam\orcidlink{0000-0001-9128-6806}}
\author{D.~Liventsev\orcidlink{0000-0003-3416-0056}}
\author{C.~Martellini\orcidlink{0000-0002-7189-8343}}
\author{A.~Martini\orcidlink{0000-0003-1161-4983}}
\author{F.~Meier\orcidlink{0000-0002-6088-0412}}
\author{S.~Mitra\orcidlink{0000-0002-1118-6344}}
\author{R.~Mizuk\orcidlink{0000-0002-2209-6969}}
\author{I.~Mostafanezhad\orcidlink{0000-0002-7111-1723}}
\author{M.~Nakao\orcidlink{0000-0001-8424-7075}}
\author{K.~Nishimura\orcidlink{0000-0001-8818-8922}}
\author{B.~Oberhof\orcidlink{0009-0002-3490-2646}}
\author{P.~Oskin\orcidlink{0000-0002-7524-0936}}
\author{P.~Pakhlov\orcidlink{0000-0001-7426-4824}}
\author{G.~Pakhlova\orcidlink{0000-0001-7518-3022}}
\author{K.~Parham\orcidlink{0000-0001-9556-2433}}
\author{A.~Passeri\orcidlink{0000-0003-4864-3411}}
\author{A.~Pathak\orcidlink{0000-0001-9861-2942}}
\author{S.~Patra\orcidlink{0000-0002-4114-1091}}
\author{I.~Peruzzi\orcidlink{0000-0001-6729-8436}}
\author{R.~Peschke\orcidlink{0000-0002-2529-8515}}
\author{M.~Piccolo\orcidlink{0000-0001-9750-0551}}
\author{L.~E.~Piilonen\orcidlink{0000-0001-6836-0748}}
\author{V.~Popov\orcidlink{0000-0003-0208-2583}}
\author{S.~Prell\orcidlink{0000-0002-0195-8005}}
\author{H.~Purwar\orcidlink{0000-0002-3876-7069}}
\author{A.~Russo\orcidlink{0009-0009-5212-8704}}
\author{D.~Sahoo\orcidlink{0000-0002-5600-9413}}
\author{S.~Schneider\orcidlink{0009-0002-5899-0353}}
\author{V.~Shebalin\orcidlink{0000-0003-1012-0957}}
\author{E.~Solovieva\orcidlink{0000-0002-5735-4059}}
\author{Z.~S.~Stottler\orcidlink{0000-0002-1898-5333}}
\author{K.~Sumisawa\orcidlink{0000-0001-7003-7210}}
\author{D.~Tagnani\orcidlink{0000-0003-0124-5088}}
\author{T.~Uglov\orcidlink{0000-0002-4944-1830}}
\author{G.~S.~Varner$^\ast$\orcidlink{0000-0002-0302-8151}}
\author{M.~Veronesi\orcidlink{0000-0002-1916-3884}}
\author{G.~Visser\orcidlink{0000-0003-2495-758X}}
\author{A.~Vossen\orcidlink{0000-0003-0983-4936}}
\author{T.~Wang\orcidlink{0009-0009-5598-6157}}
\author{X.~L.~Wang\orcidlink{0000-0001-5805-1255}}
\author{L.~Wood\orcidlink{0000-0002-5235-8181}}
\author{X.~P.~Xu\orcidlink{0000-0001-5096-1182}}
\author{K.~Yoshihara\orcidlink{0000-0002-3656-2326}}
\author{Y.~Zhai\orcidlink{0000-0001-7207-5122}}
\author{V.~I.~Zhukova\orcidlink{0000-0002-8253-641X}}

\begin{abstract}
The K-long and muon detector (KLM) constitutes the outer-most volume of the Belle~II spectrometer at the interaction region of the SuperKEKB collider in Tsukuba, Japan.
The KLM detector was partially upgraded since the Belle experiment by replacing many of its resistive-plate chambers with scintillators containing wavelength-shifting fibers and instrumenting it with silicon photomultipliers.
We describe the readout electronics, firmware, and software created to control and acquire data from the scintillators and resistive-plate chambers.
\end{abstract}



%
%
%

\end{frontmatter}


\customfootnotetext{*}{Author deceased at time of publication}

\section{Introduction} 
The K-long ($K_L^0$) and muon ($\mu$) detector, or KLM\footnote{
  Common, non-standard acronyms: K-long and muon (KLM); TeV Array Readout with GSa/s Sampling and Event Trigger (TARGETX) where X denotes the production version; Standard Control and Read Out Device (SCROD)
}
, makes up the outermost active
volume of the Belle~II detector, located at the interaction point of the SuperKEKB $e^+e^-$ particle 
collider in Tsukuba, Japan. The Belle~II detector is a roughly 3-story-tall general-purpose particle spectrometer
designed to detect particles with energies between about 50 and \SI{7000}{\mega\electronvolt}.
The flux return of its \SI{1.5}{\tesla} solenoidal magnetic field consists of 14 layers of \SI{4.7}{\cm}-thick steel plates.
These plates form octagons in the Belle~II barrel region and flat disks in the endcap regions.
The gaps between the plates are interleaved with active particle detection modules: 
15 layers in the barrel, 14 in the forward endcap, and 12 in the backward endcap.
Besides serving as the flux return, the steel plates also offer more stopping power for hadrons, contributing 
an additional 3.9 interaction lengths in addition to the 0.8 interaction lengths of the electromagnetic calorimeter\cite{abe2010belle}.

During the first-generation Belle experiment\cite{ABASHIAN2002117} (1999-2010), the KLM detector was instrumented exclusively with resistive-plate counters (RPCs)\cite{Wang2003RPCPA}.
Because of the expected high neutron background when Belle~II is operating at 
its design luminosity, the inner two barrel layers and all of the endcap layers
have been replaced with plastic scintillators.
The chosen scintillators are long strips with a \SI{1 x 4}{\cm}
cross section in the barrel and \SI{0.7 x 4}{\cm} in the endcaps. Strip lengths vary due to the geometry of the detector.
Each scintillating strip contains a multi-clad \SI{1.2}{\mm} diameter wavelength
shifting fiber running down its central axis, and a silicon photo-multiplier (SiPM) at 
one end of the fiber. Details of the scintillator and wavelength-shifting fiber selection and construction can be found in Ref.~\cite{AUSHEV2015134}.

This paper describes the KLM electronic readout system.
We discuss the RPC and the scintillator readout systems separately up to 
the point where the two data streams are merged.
We organize the sections according to the flow of information, starting with a charged particle
passing through a single RPC module or KLM scintillator bar and following the signal
through the data acquisition system.
Details about the calibration of the SiPMs and the TARGETX (``an oscilloscope on a chip") form the subject of the 
final section.

\section{Scintillator Readout Electronics}
Charged particles passing through the KLM generate scintillation light in the scintillator strips. 
Some of the light in each strip is absorbed by its central wavelength-shifting fiber.
The wavelength shifter is a material that absorbs light at a higher frequency and reemits it at a lower frequency,
and it has a longer attenuation length than the plastic scintillator material. 
As the wavelength shifter reemits this light isotropically, photons
emitted at angles less than the critical angle of the fiber
propagate to the ends of the fiber.
When a photon hits one of the 667 pixels of the SiPM\footnote{Hamamatsu s10362-13-050c} (each pixel is an avalanche photodiode),
there is a roughly \SI{20}{\percent} possibility that an avalanche forms. In an avalanche,
one photoelectron is rapidly amplified to about 750,000 electrons. 
Multiple photons can fire multiple pixels and their outputs are combined in parallel.
Current across quenching resistors in the SiPM decreases the bias voltage
across the photodiode and the avalanche(s) cease. The result is a signal 
with a steep leading edge, a long tail, and an amplitude proportional 
to the number of pixels that fired.

\begin{figure}
  \centering\includegraphics[width=\columnwidth]{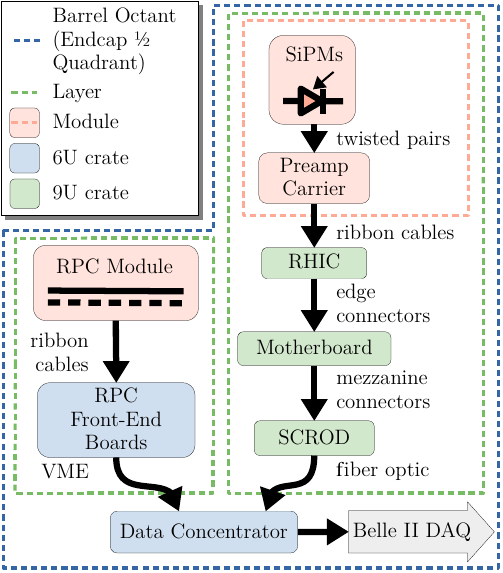}
\caption{Flow diagram of KLM readout. For each layer of each octant (8 forward and 8 backward), there
is one RPC or scintillator module. In each layer of each endcap quadrant, there are only scintillator modules.
Each RPC module connects via ribbon cables to an RPC front-end board in
a 6U VME crate, while each scintillator module connects via ribbon cables to one RHIC-and-motherboard combination
in a 9U crate. In a barrel (endcap) scintillator module, there are 7 (10) preamplifier carriers 
and about 78 (exactly 150) SiPMs in each. In the barrel (endcap), 15 (7) front-end boards connect to one Data Concentrator.}\label{fig:eklm_ro_topo}
\end{figure}

The following sections describe the electronics encountered by a signal
on its way to the Belle~II data acquisition (DAQ) system\,\cite{B2DAQ}. Figure \ref{fig:eklm_ro_topo}
depicts the multiplicity of each component in the KLM readout and how each is connected.

\subsection{Preamplifiers}
In each scintillator module, groups of 15 SiPMs are connected to
a preamplifier carrier card via twisted-pair cables, and, in turn, each carrier card is 
connected to the readout and control electronics located at the top of the Belle~II magnet yoke via two ribbon cables several
meters in length. One ribbon cable supplies a unique bias voltage to each SiPM, while the 
other delivers each preamplified signal to the readout system. 

We use a custom-designed preamplifier that is a fully-differential operational amplifier. 
Each is assembled on a \SI{2}{cm} by \SI{2}{cm} printed circuit board that plugs edgewise into the preamplifier carrier card.
The preamplifier gain is large enough to resolve single-photoelectron pulses from a SiPM.
This was an oversight, perhaps, as even minimum-ionizing particles traversing the KLM detector tend to saturate the preamplifier. 
While this large preamplifier gain enables measurement of SiPM gain using single-photon spectra, which we describe later, the saturation prevents us from resolving the full pulse height of some hits during Belle~II operation.
Further, we cannot measure the leading-edge time of a pulse by using a dynamic threshold set relative to the pulse height.
Rather, we opt to measure leading-edge time using a fixed threshold.
The preamplifier input from a SiPM is indicated in Fig.~\ref{fig:SiPM_circuit} in the following section, but a schematic of the preamplifier circuit itself is not shown.

\subsection{Ribbon Header Interface Card}
All of the ribbon cables from a single scintillator module are connected to a single
set of readout and control electronics via the Ribbon Header Interface Card (RHIC). This
circuit board connects all of the preamplified signals to the scintillator motherboard,
and it also has two octal 8-bit \SI{5}{\V} digital-to-analog converters\footnote{
Texas Instruments DAC088S085
} (DACs) for each group of 15 SiPMs to fine-tune 
the SiPM bias voltage with a precision of \SI{20}{\mV}.
We refer to these as the HV-trim DACs.  

\begin{figure*}
  \centering\includegraphics[width=4.75in]{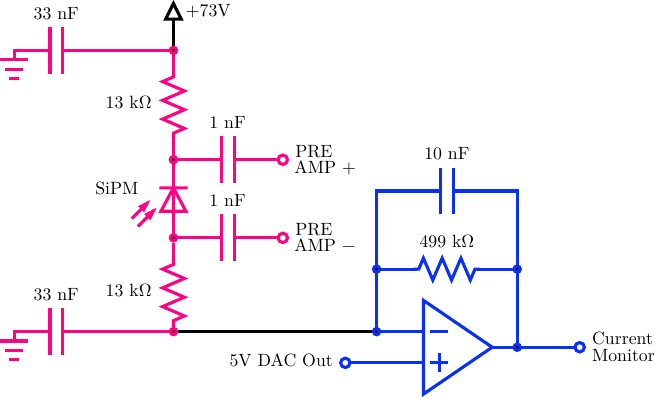}
  \caption{Simplified schematic of SiPM bias and current monitor. Pink components, including
  the preamplifier itself, are located
  inside a scintillator module and blue components are part of the RHIC. The \SI{5}{V} DAC output
  is used to create a virtual termination for the SiPM via the inputs of a charge-sensitive amplifier.
  This allows fine tuning of the gain in \SI{20}{\mV} steps.
  The current monitors of all channels are connected to the Standard Control and Read-Out Device (SCROD) via multiplexers and an analog-to-digital converter. 
  Their purpose is to monitor dark current of each channel over the life of the experiment. SiPM output is
  captured by the inputs of the preamplifier prior to being transmitted over the ribbon cables.}
  \label{fig:SiPM_circuit}
\end{figure*}

Each DAC output is buffered through a charge-sensitive amplifier\footnote{
Texas Instruments LMV324
}
--- the DAC drives the ($-$) terminal, the ($+$) terminal terminates the SiPM, and the amplifier output serves as a current monitor (Fig.~\ref{fig:SiPM_circuit}).
Each RHIC also contains a 2-pin high-voltage (HV) connector and 
distributes the HV to all preamplifier carriers over the ribbon cables. 
\subsection{Scintillator Motherboard}
\begin{figure}
  \centering
  \includegraphics[width=0.49\columnwidth]{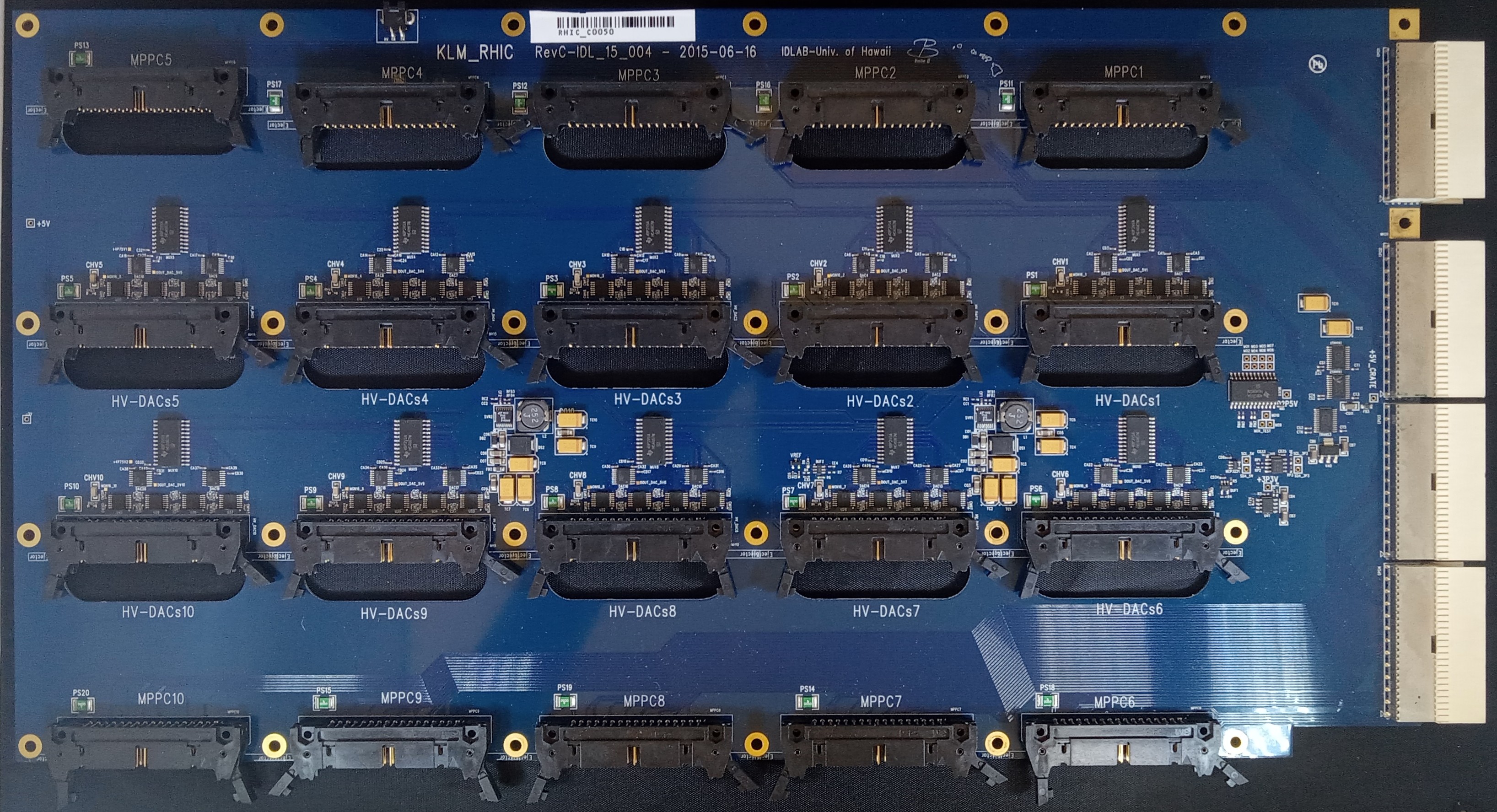}
  \includegraphics[width=0.49\columnwidth]{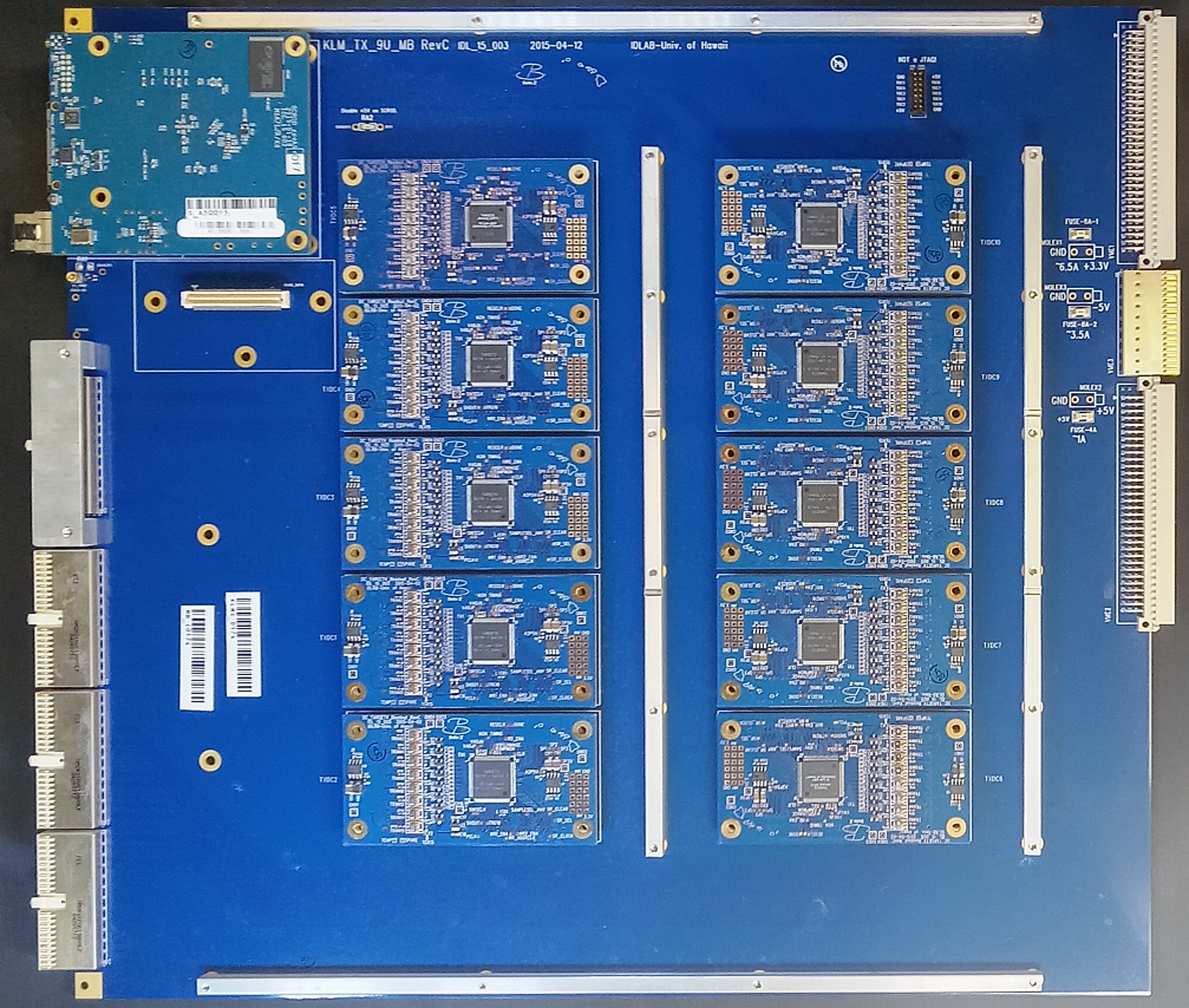}
  \caption{Left: ribbon header interface card (RHIC).
  Right: scintillator motherboard.
  The RHIC Provides voltage to SiPMs and collects signals.
  It contains 20 octal DACs for setting the voltage.
  They are 8-bit \SI{5}{\V} DACs and are used to fine tune the ground side of the SiPM bias.
  The scintillator motherboard has 10 TARGETX daughter cards and a SCROD mounted to it.}
  \label{fig:RHIC_and_MB}
\end{figure}

Each RHIC is connected edge-to-edge with a scintillator motherboard that is installed in a 9U Versa Module Eurocard (VME) crate (Fig.~\ref{fig:RHIC_and_MB}).
The VME crate is only used for low-voltage supply to the scintillator readout electronics.
The scintillator motherboard has 10 TARGETX daughter cards 
(waveform-sampling ASICs described in the next section)
connected to its top. Each group of 15 SiPMs is
routed to one daughter card. The motherboard is arranged in a two-bus configuration, 
each with independent control circuits and each with 15 data lines. As the largest scintillator
modules (located in the endcaps) contain 75 vertical ($x$) strips and 75 horizontal ($y$) strips, this two-bus configuration allows
for the readout of $x$ hits and $y$ hits simultaneously. These buses are routed to a single
Standard Control and Read-Out Device (SCROD) board.

\subsection{TARGETX Daughter Card and the TARGETX}
\begin{figure*}
  \centering
      \includegraphics[height=1.5in]{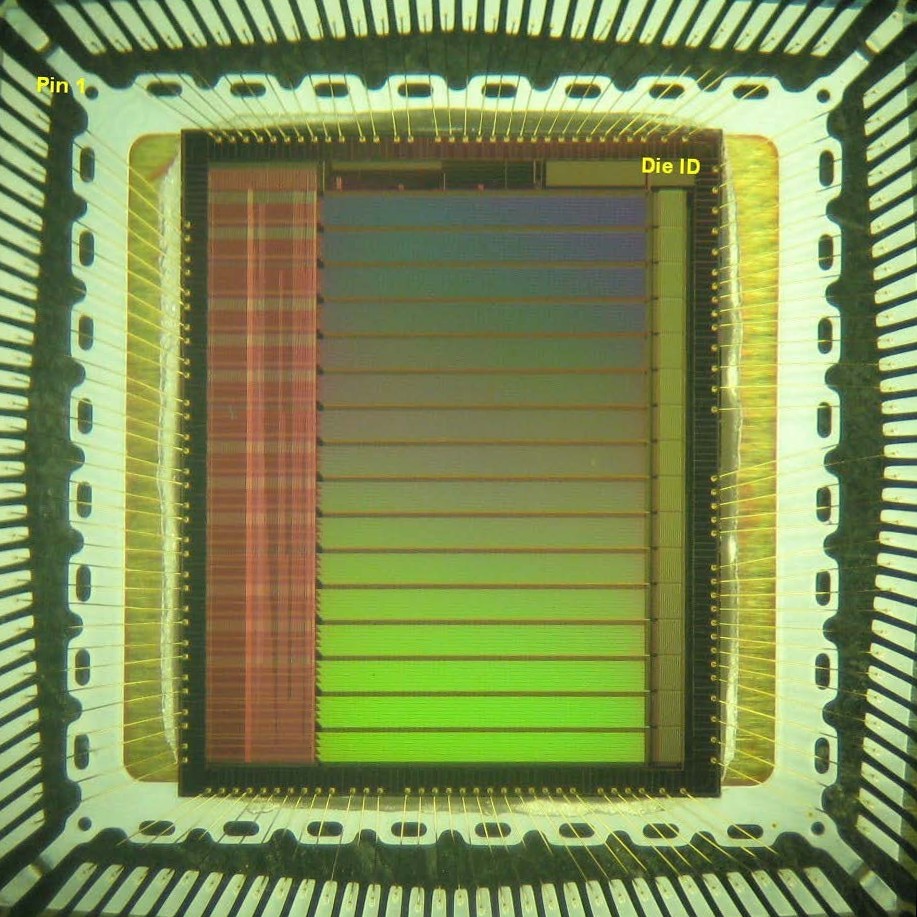}
      \includegraphics[height=1.5in]{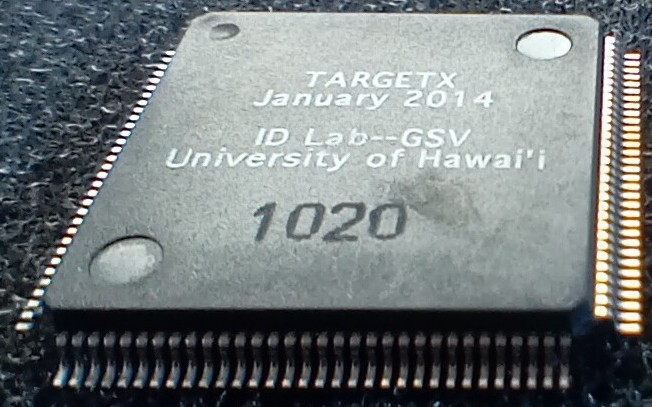}
      \includegraphics[height=1.5in]{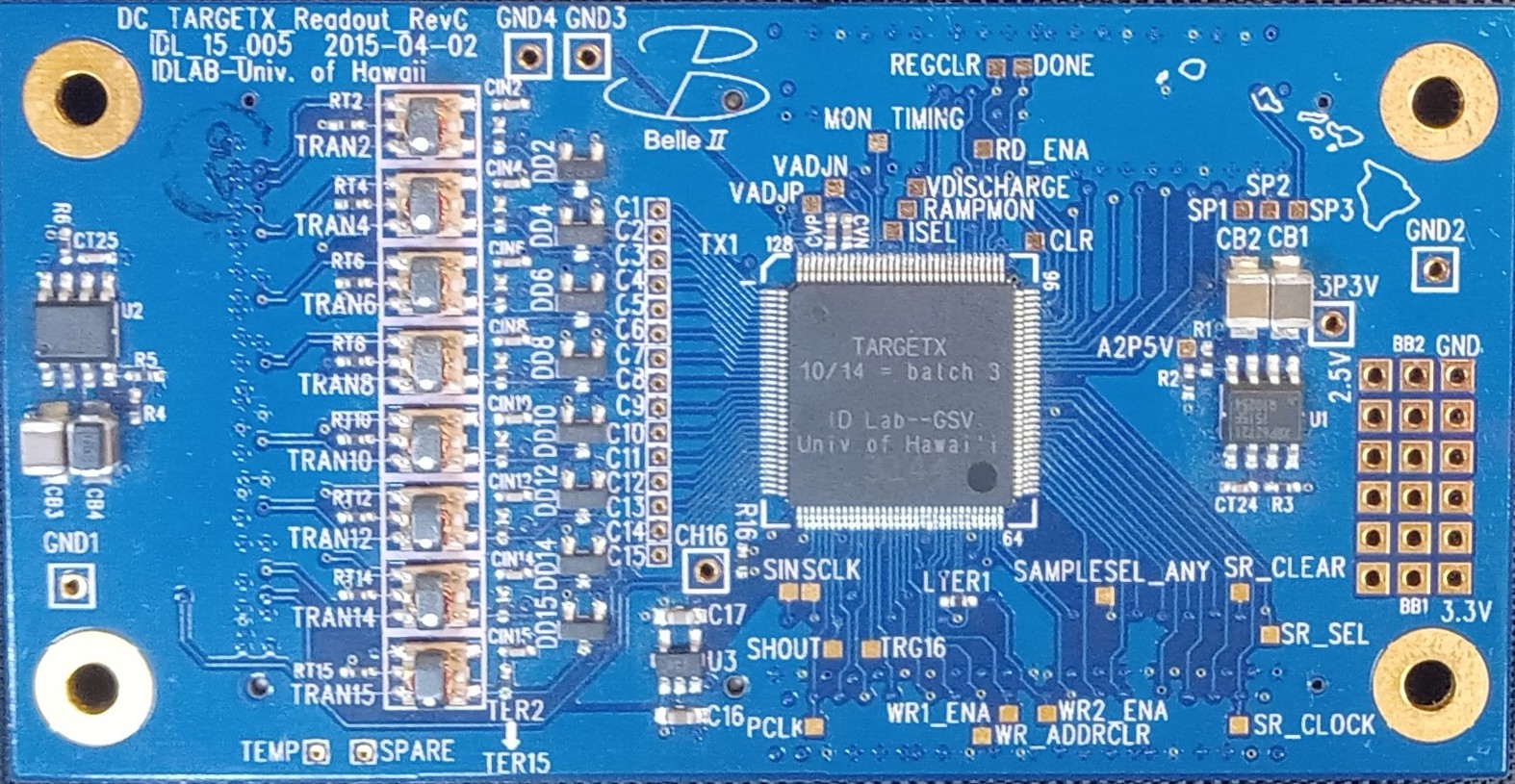}      
  \caption{TARGETX die photograph, TARGETX in its package, and TARGETX Daughter Card.}
  \label{fig:TARGETX_and_DC}
\end{figure*}
The TARGETX (Fig.~\ref{fig:TARGETX_and_DC}) is a waveform-sampling application-specific integrated circuit (ASIC). The TARGETX daughter card AC-couples the SiPM signals via transformers and provides over- and under-voltage protection
for the TARGETX inputs. Aside from containing various probe points for testing, its sole purpose is to house a single TARGETX ASIC. 

The TARGET (TeV Array Readout with GSa/s sampling and Event Trigger)
series of ASICs was originally designed for the readout of Cherenkov cameras\,\cite{Vandenbroucke_2012}.
The TARGETX is the latest version of TARGET and was designed specifically for KLM scintillator-based readout to include triggering capabilities. 
It is a 16-channel\footnote{
KLM only uses 15 TARGETX channels. The 16th channel of each ASIC is used for bench testing.}
analog storage device, capable of sampling at \SI{1}{\GHz} using $2^{14}$ sample-storage cells per channel, 
allowing it to store \SI{16.384}{\micro\second} of analog data. 
Every channel has two switched-capacitor sampling arrays of 32 sampling capacitors each; while one array is sampling the other is being transferred
to a capacitor-based analog-storage array.
Each channel's storage array is composed of 512 \textit{windows} of 32 storage capacitors each.
Storage takes place in a round-robin fashion,
always moving the write-address pointer sequentially across the storage windows.
The write pointer can
be synchronized with the user's application by asserting a clear signal thus resetting the write address to zero.
To prevent overwriting, the user can deassert the write-enable signal as the write pointer
moves across some region of interest, but this will cause some dead time
as incoming samples will have nowhere to be written.
Each channel also has a fast trigger output 
with a programmable 12-bit trigger threshold. Finally, for digitization, the TARGETX uses a Wilkinson 
analog-to-digital converter (ADC). 
There is one Wilkinson ramp generator and 32 fast Gray-code counters per channel allowing all 32 samples of
a selected storage window to be digitized simultaneously on all 16 channels.
The device has one data-out pin per channel and a 14-bit address select bus;
hence, data from all of its channels can be shifted out in parallel, starting and stopping
on any of the $2^{14}$ storage cells desired.
Refs.~\cite{Vandenbroucke_2012, Tibaldo_2015, AIP_2016, Vandenbroucke_2017} provide more information about the TARGET series of ASICs.

\subsection{Standard Control and Read Out Device (SCROD) Board}
\begin{figure}
  \centering
  \includegraphics[width=0.7\columnwidth]{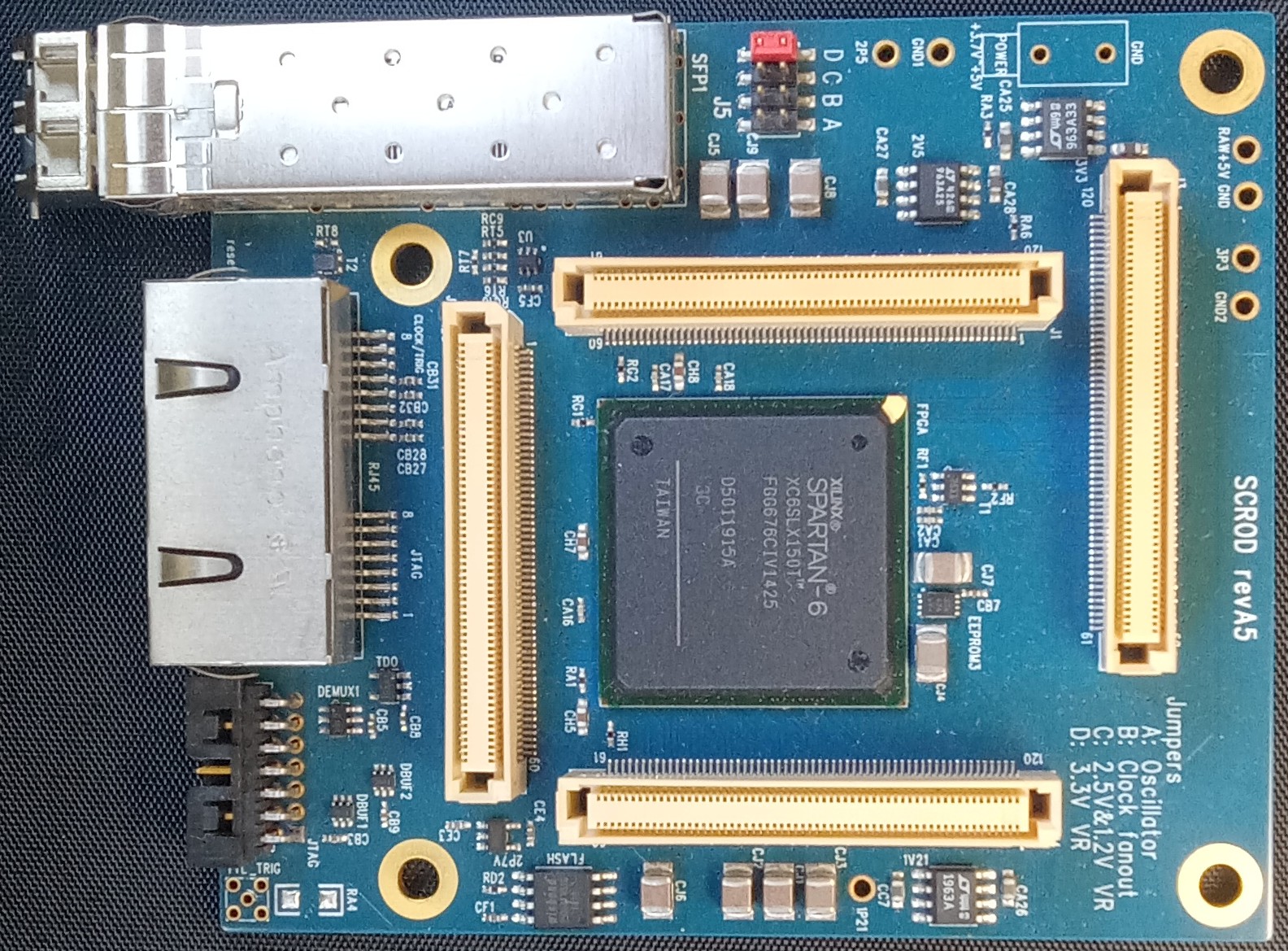}
  \caption{A SCROD board. This board is basically the command center for the scintillator readout electronics. It controls up to 10 TARGETX ASICs all the while listening for triggers and command words, processing waveform data, and delivering data packets.}
  \label{fig:SCROD}
\end{figure}
Also connected to the top of the motherboard is the final
piece of dedicated readout electronics for the scintillator system, the SCROD (Fig.~\ref{fig:SCROD}).
This board contains one field-programmable gate array (FPGA)\footnote{Xilinx Spartan-6 XLS-150T} and serves as the interface to the rest of the Belle~II readout system. Global (detector-wide)
synchronous clock and global trigger arrive via low-voltage differential signals over network
cables to one registered-jack 45 (RJ45) connector on the SCROD. A second
RJ45 brings in the JTAG (joint-test action group) signals needed for reprogramming the FPGA.
Lastly, a serial
fiber transceiver provides two-way communication with the rest of the detector's readout system.
It can operate using either an external clock source or its onboard oscillator by 
the addition of a jumper resistor. A static random-access memory chip (SRAM)\footnote{
Infineon Technologies CY62177EV30
}, which is used
in its 4\,M by 8-bit configuration, provides additional storage for the FPGA.

\section{Scintillator Readout Firmware}
Control of 10 TARGETX ASICs, waveform readout and processing, L0 trigger buffering, and
L1 trigger processing are all managed by a single SCROD (L0 refers to channel self triggers from the TARGETX while L1 
refers to triggers from the Belle~II global decision logic, basically a request
to the front-end electronics to check their buffers and report any hits within
the time window of interest). 

\begin{figure*}
  \centering
  \includegraphics[width=6.4in]{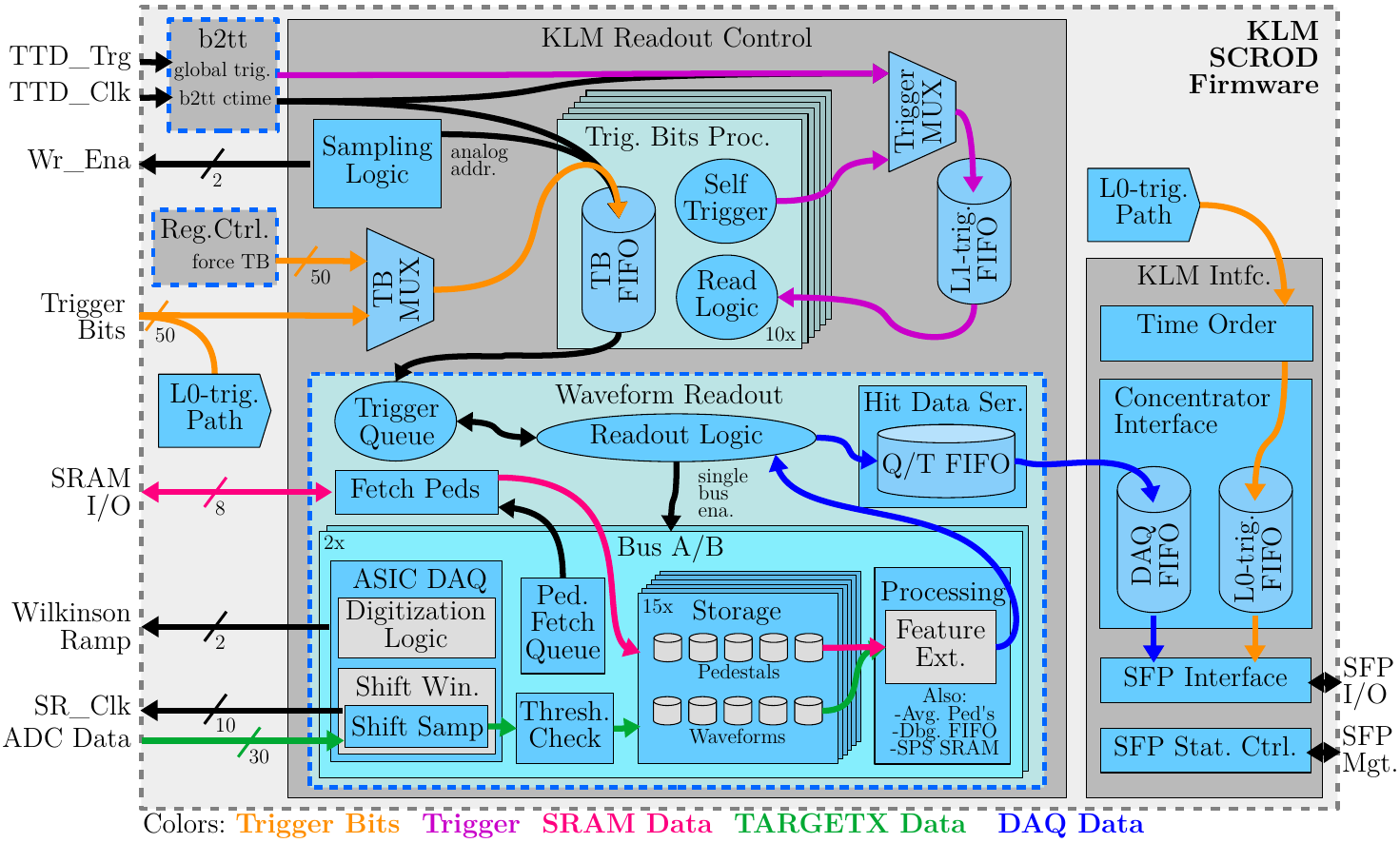}
  \caption{Block diagram depicting the layout of the SCROD firmware. The core entity of the waveform 
  readout logic is duplicated twice, once for each bus of five TARGETX ASICs. Similarly, the storage 
  entity is duplicated 15 times on each bus since all 15 channels of a TARGETX can be read out simultaneously}
  \label{fig:SCROD_block_diagram}
\end{figure*}

The firmware can be described by breaking it into two principal paths through which data flows.
These two main paths are the L0 trigger path and the digital
waveform path.  In Fig.~\ref{fig:SCROD_block_diagram}, these two paths are colored orange and green, respectively.
For brevity, peripheral processes like the configuration and status register interface,
phase-locked global clock input, fiber transceiver interface, and SRAM access are not discussed.

\subsection{L0 Trigger Path}
In the TARGETX, the analog input of each channel is tied to one input of a comparator.
The comparator reference input is the programmable trigger threshold. When the analog 
input crosses the threshold, the comparator triggers a fixed-length digital pulse
generated by a one-shot timer. 
One-shots from all 15 channels are fed into an
address encoder, and finally delivered to the FPGA using five \textit{trigger} bits.
Four bits are sufficient to encode 15 channels by reporting the channel number in binary.
Sixteen-channel encoding is not possible because zero is reserved for the
case when there are no triggers. The fifth (most significant) trigger bit handles the case when
more than one channel fires simultaneously on a single TARGETX. In this case, trigger
bit 5 is held high and the first four (least significant) bits encode which group of 4(3) channels may have been hit (the last group contains 3 channels).
For example, the five-bit pattern 0b10001 means multiple channels in the first group (channels 1, 2, 3, and 4) have been hit,
while the five-bit pattern 0b11100 means multiple channels in the last two groups (channels 9 through 15) have been hit.
We refer to these as multi-channel hits. For such hits, waveform digitization is required to 
disambiguate which channels in the channel groups actually had hits.
If waveform digitization is not used, the detector resolution is degraded
from \SIrange{4}{16}{\cm}, and the number of strips hit in each group is unknown.

When L0 triggers from any of the TARGETX ASICs on the motherboard arrive at the FPGA, they
are copied, timestamped, and split into two paths. One path performs time ordering
of the hits and sends the time-ordered hit information via fiber optics to the 
Data Concentrator. These hits are transmitted to the 
Belle~II trigger decision logic. The other path timestamps the bits again,
this time using the TARGETX write address, and writes them to a FIFO (first-in first-out), where they wait to
either be matched to an L1 trigger and go on to digitization, or to exceed the maximum look-back time and be cleared from the FIFO.
There is one such FIFO allocated for each of the 10 TARGETX ASICs.

\subsection{L1 Trigger Handling}
After an L0 trigger is sent, the Belle~II global trigger
logic calculates (based on L0 triggers from all of the subdetectors) whether or
not a global (L1) trigger should be issued. The latency of this decision is fixed, so when
a trigger is issued, every subdetector only need look back a finite amount of time
(less than \SI{5}{\micro\second}) to see if there were any corresponding L0 triggers within that time frame. 

When an L1 trigger is received, the SCROD firmware quickly checks its hit buffer
and earmarks any hits for digitization. 
The first step is to set up a mask over the region of interest to prevent the TARGETX
from overwriting the analog storage cells before digitization can finish. If 
more L1 triggers are received while a previous one is being digitized, they
are also masked immediately and earmarked for later digitization. 
To prevent event pileup, the digitization queue has a programmable threshold
at which it will forego digitization in order to catch up. When this threshold
is reached and a previous L1 trigger has just finished processing, the remaining
jobs in the queue are processed in \textit{simple} mode, wherein hit time is just
the timestamp of the L0 trigger, pulse height is reported as zero, and one extra
bit within the data packet is toggled to indicate that waveform digitization was
not carried out for that hit.

Simulation tests verify that this scheme allows the SCROD firmware
to keep pace with a \SI{30}{\kHz} L1 trigger rate (Poisson-distributed with a minimum of  \SI{200}{\ns} between consecutive L1 triggers),
the requirement for Belle~II unified readout-system design\,\cite{Nakao2021}.
Analysis of physics data taken at a luminosity of \SI{1.9e34}{\cm^{-2}\s^{-1}}
and with an L1 trigger rate of \SI{2.8}{\kHz} shows that digitization is skipped
for only \SI{0.4}{\percent} of hits.

\subsection{Digitization}
To fetch waveforms from the TARGETX analog memory, the SCROD firmware 
asserts a 9-bit window address to each and 
enables a Wilkinson ramp on one or more of them. The 32 samples in a given storage window
are digitized simultaneously. The time required to digitize depends on the 
slope of the Wilkinson ramp, which is tunable and can range from less than
one microsecond to many tens of microseconds. 
Tuning of the Wilkinson ramp is described in Section~\ref{section:ramp_tuning}.

When ADC conversion is finished, the SCROD firmware supplies a shift-register clock to one TARGETX and
reads back all 15 channels in parallel, one bit at a time. While all 10
TARGETX ASICs can be digitized simultaneously, data can only be shifted out from
two of them at a time (one on each bus). Of the 32 samples digitized for a given window,
a 5-bit sample select signal is asserted to choose which sample to shift out. Once
a sample is shifted from an ASIC to the SCROD, it is written to a FIFO (\textit{Waveform FIFOs}).
A total of 150 FIFOs (each 12-bit wide and 512-bit deep) are allocated for 
this purpose --- one for each channel on the motherboard.

\begin{figure}
  \centering
  \includegraphics[width=\columnwidth]{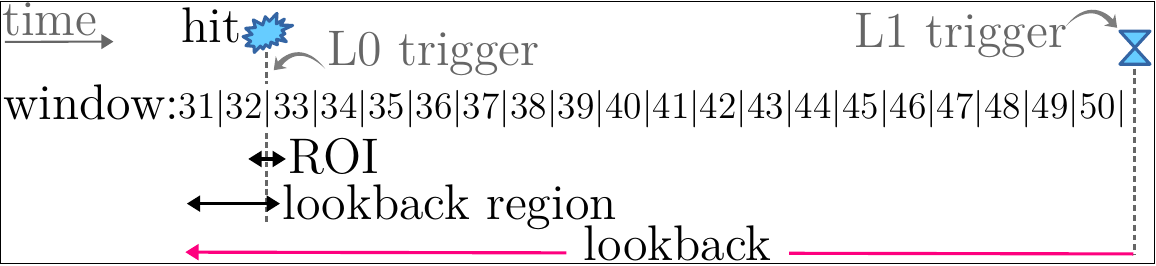}
  \caption{Depiction of lookback time, region of interest, and TARGETX storage windows. In this example, 
  the initial hit is timestamped (L0 trigger) in the firmware at the end of 
  storage window number 32. After an L1 trigger is issued, the lookback region is calculated and the hit buffer is checked for hits.
  Since there is a hit contained in the lookback region, the region of interest is calculated, and in this case, it spans the boundary of windows 32 and 33, so both windows must be digitized.}
  \label{fig:lookback}
\end{figure}
The region of interest (ROI) is calculated by the SCROD firmware in advance based on the
timestamp of the L0 trigger. It may be contained in one storage window or span
several storage windows. In the latter case, as in Fig.~\ref{fig:lookback}, each window must be digitized
and shifted out in sequence. The use of one FIFO per channel avoids complications that would arise from 
multiplexing data across multiple channels and ASICs. 

A register in the SCROD firmware sets the number of samples
to read out. The minimum number of samples is four, limited by the ROI calculation
logic, and the maximum is 512, limited by the depth of the waveform FIFOs.
Another register sets a sample-number offset relative to the 
L0 timestamp. This allows fine-tuning of the \textit{lookback} time --- just as
one would turn the horizontal-translation knob on an oscilloscope to center
their signal in the screen.
 
Before analyzing the waveforms, they must be cleaned up. Every storage cell
in the TARGETX has a slightly different DC offset, known as the pedestal voltage.
The pedestal voltage for each storage cell must be measured beforehand 
in a calibration sequence so that it can later be
subtracted from the waveform. Another 150 FIFOs (also 12-bit by 512-bit) are
allocated for the waveform pedestal subtractions.

\subsection{Pedestal Management}
Pedestals are stored on the SCROD's 4\,M$\times$8-bit static RAM (SRAM) chip.
Whether reading or writing, SRAM access takes \SI{55}{\ns} per address. Since TARGETX
samples have a 12-bit resolution, pedestal values for two sample cells are stored over three
SRAM addresses. 
On the FPGA, the logic for reading and writing the waveform and pedestal FIFOs contains
a normal (data acquisition) mode and a measurement mode. 

In the normal mode, while the region of interest is being digitized and shifted out by the digitization logic,
another firmware entity
begins fetching pedestal values from the SRAM and writing them to the pedestal FIFOs.
As there is only one SRAM to store pedestals for all ten TARGETX ASICs, pedestal 
reading necessarily happens sequentially.
If a TARGETX ASIC experiences a multi-channel hit, 
pedestals must be fetched for every channel that may have been hit (between
3 and 15 channels, depending on the trigger-bit pattern on the multi-channel hit). 
Pedestal reading generally takes less time than shifting out waveforms from the 
TARGETX ASICs, but in the case of a multi-channel hit with a trigger bit pattern
requiring several channels to be checked, e.g.~0b11111, SRAM access becomes the
bottleneck. 

The firmware entity responsible
for pedestal reading has a fair arbiter that prioritizes SRAM scheduling for one
channel on one bus, and then one channel on the opposite bus. It remembers which bus
was serviced last and services the opposite bus next if arbitration between
the two buses is required. This feature was added to prevent biasing one
bus over the other in case a make-haste signal is applied and some
hits in an event have to forego both digitization and/or feature extraction.
Such a signal is not implemented at this time but may be required in the future when SuperKEKB luminosity,
detector background rates, and L1 trigger rates all increase.

\begin{figure*}
  \centering
  \includegraphics[width=5.0in]{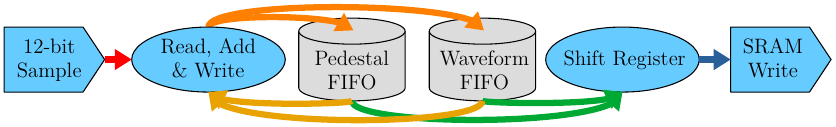}
  \caption{Resource-saving scheme to use existing FIFOs for measuring pedestals. 
           Concatenating inputs and outputs of two 12-bit FIFOs allows pedestals to be measured up to $2^{12}$ times. A shift register performs the averaging.}
  \label{fig:ped_meas}
\end{figure*}
In measurement mode, the firmware hijacks the digitization machinery described
above, the voltage of each storage cell is measured many times (up to $2^{12}$), and
the average pedestal values are written to the SRAM. This must be done
in advance, with the HV turned off, so that the values are available during data taking.
In this configuration (Fig.~\ref{fig:ped_meas}), inputs and outputs of one waveform FIFO and one pedestal
FIFO are concatenated to make a single 24-bit wide FIFO. Prior to pedestal
measurement, all the FIFOs are primed with 32 writes of the value zero (the number of
samples in a TARGETX storage window). During
pedestal measurement, a command to write to the FIFOs causes them to first be
read once, their 24-bit output is added to the 12-bit sample, and the sum is written back 
into the FIFOs. 
It is repeated $2^\mathcal{N}$ times for each of the 512 storage windows, where $\mathcal{N}$ is set
by a SCROD register. Averaging is achieved by shifting the 24-bit result to
the right $\mathcal{N}$ times and then keeping the 12 least-significant bits.

\subsection{Feature Extraction}
While processing an L1 trigger, the feature extraction entity is enabled
as soon as any of the channels have both their associated waveform and pedestal FIFOs ready.
The two FIFOs are read in tandem and the pedestals are subtracted from the waveform samples.
To avoid the use of negative numbers, a baseline value of 3072 (3/4 full scale) is added
to each sample during pedestal subtraction.

\begin{figure}
  \centering
  \includegraphics[width=\columnwidth]{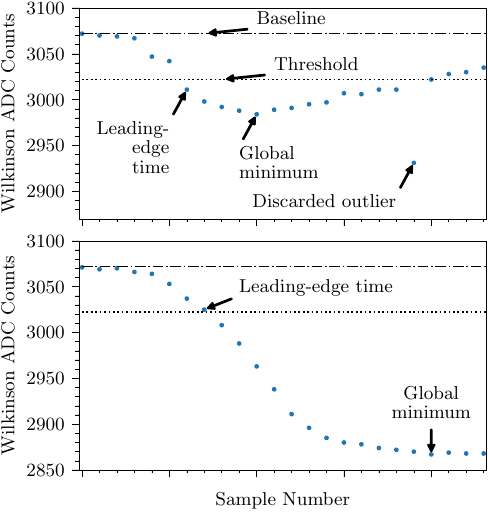}
  \caption{Example waveforms showing the SCROD firmware's feature-extraction algorithm.
  Top: Normal pulse.
  The sample closest to the discrimination threshold is recorded as leading-edge time. 
  The baseline minus the global minimum is recorded as pulse height.
  The outlier is rejected.
  Bottom: Pulse that saturated its preamplifier as can be seen by the plateau shape.}
  \label{fig:feat_ext}
\end{figure}

Leading-edge time is measured using constant-value discrimination with linear
interpolation to find the sample nearest the threshold. Constant
fraction discrimination is not used due to the saturation of
the preamplifiers for very large pulses (Fig.~\ref{fig:feat_ext}). Pulse height is measured by finding the
global minimum (pulses are negative going), requiring equal or larger samples on either side, 
and recording its magnitude with respect to the baseline. An outlier monitor discards any potential minimum sample that
is further than 128 ADC counts from the average of its two neighbor samples.
Outliers may occur if there is a bit error when shifting a sample out of the TARGETX.
If a small pulse does not cross the discrimination threshold, the 
time of the minimum is substituted for the leading-edge time. 
If a global minimum is not found (waveform is constantly increasing or decreasing),
then the last sample is used for the pulse height.
Time and pulse height are written to a data packet,
and sent over a Xilinx Aurora data link to the Data Concentrator.

At the same time (for calibration purposes) the waveform is written to a debugging FIFO, and the measured pulse height
is written to a histogram, both of which can be read back
through the register interface.
Three debugging FIFO modes are available: write the waveform with pedestal subtraction,
write the waveform without pedestal subtraction, and write the pedestals only.

Once all valid channels in the event have been processed, the remaining
waveform FIFOs corresponding to channels that did not have valid hits but were
shifted out of the TARGETX anyway are all read until they are empty, so that
the firmware is ready to process the next event in the queue.

\section{RPC Front-End Board}
RPC signals are generated when a charged particle leaves an ionization trail in the gas gap of an 
RPC module. The strong electric field creates an avalanche, and electric current is induced
on the two orthogonal readout-strip planes in the vicinity of the avalanche. Like the scintillator readout, the RPC readout
relies on ribbon cables to transport the signal to the readout electronics. 
Unlike the scintillator readout, the RPC readout is not instrumented with 
waveform digitizers. It discriminates the signal and timestamps it before
transmitting the timestamp to the Data Concentrator.
\begin{figure*}
  \centering
  \includegraphics[width=0.7\linewidth]{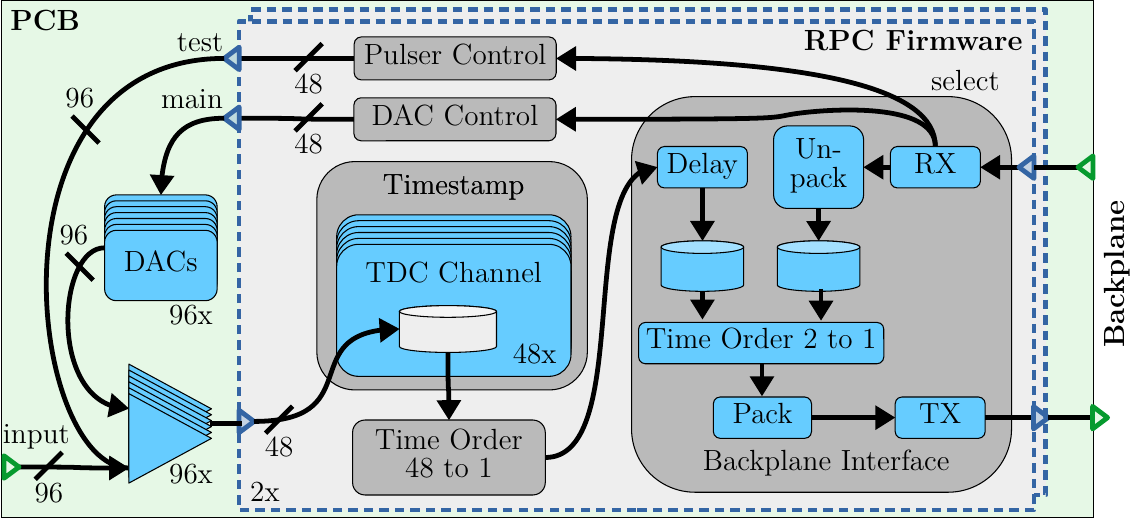}
  \caption{Schematic of an RPC front-end board and its firmware. 
           It has two FPGAs, each one handling 48 RPC channels.
           Signals from RPC channels that exceed the programmable
           digitization threshold are timestamped and transmitted
           over the VME backplane to the Data Concentrator.
           There are 13 RPC front-end boards per Data Concentrator
           in the barrel region of the KLM detector.}
  \label{fig:RPC_FEB_fw}
\end{figure*}

A hit is formed when an electric pulse from an RPC readout strip exceeds a discrimination threshold set by a DAC\footnote{Analog Devices LTC2636} on the RPC front-end board. 
The basic layout of the RPC front-end board and its firmware is depicted in Fig.~\ref{fig:RPC_FEB_fw}.
In the RPC front-end firmware, hits are timestamped in a time-to-digital conversion (TDC) module with a resolution of \SI{3.94}{\ns} (2$\times$system clock).
Each RPC front-end board contains 96 line receivers and discriminator channels, 48 per front-panel (ribbon cable) connector. 
Channels 1-48 connect to negative RPC pulses while channels 49-96 
connect to positive RPC pulses.
An analog test pulser provides an independent built-in test of each channel.
Two FPGAs\footnote{Xilinx Spartan-6 XC6SLX25} are used for discriminator control and timestamp generation.
These FPGAs are configured over the backplane using SERA/A08/A09.
The threshold and pulser are operated
with run control commands over the Belle2Link.
The discriminators only generate a rising edge for the FPGA timestamp generator.
Hits are also time ordered, using their timestamps, to simplify event building in the Data Concentrator.

\begin{figure}
  \centering
  \includegraphics[width=0.8\columnwidth]{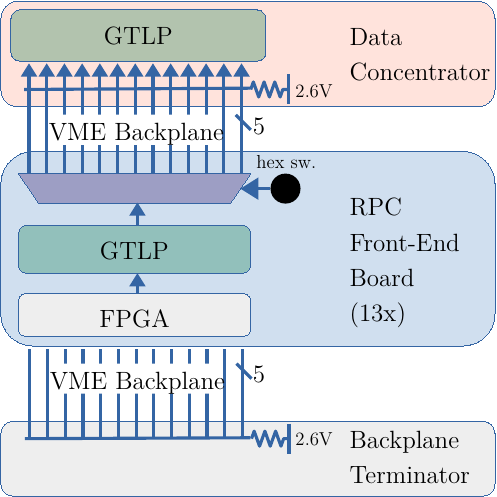}
  \caption{Custom backplane protocol using rotary switches to 
           address a demultiplexer and communication over 5-bit
           buses using GTLP.}
  \label{fig:RPC_backplane}
\end{figure}

Hits are transmitted to the Data Concentrator over the VME backplane using a custom protocol.
The backplane termination voltage is lowered to \SI{2.6}{\V} for GTLP (Gunning Transceiver Logic Plus).
TDC data is transmitted over dedicated 5-bit buses.
Each 5-bit bus is 1-13 demultiplexed to the desired
slot/position which corresponds to its layer.
The position is selected by a hex rotary switch on the RPC front-end board 
which must be set during installation (Fig.~\ref{fig:RPC_backplane}).

\section{The Data Concentrator}
\begin{figure}
  \centering
  \includegraphics[angle=90, width=0.49\columnwidth]{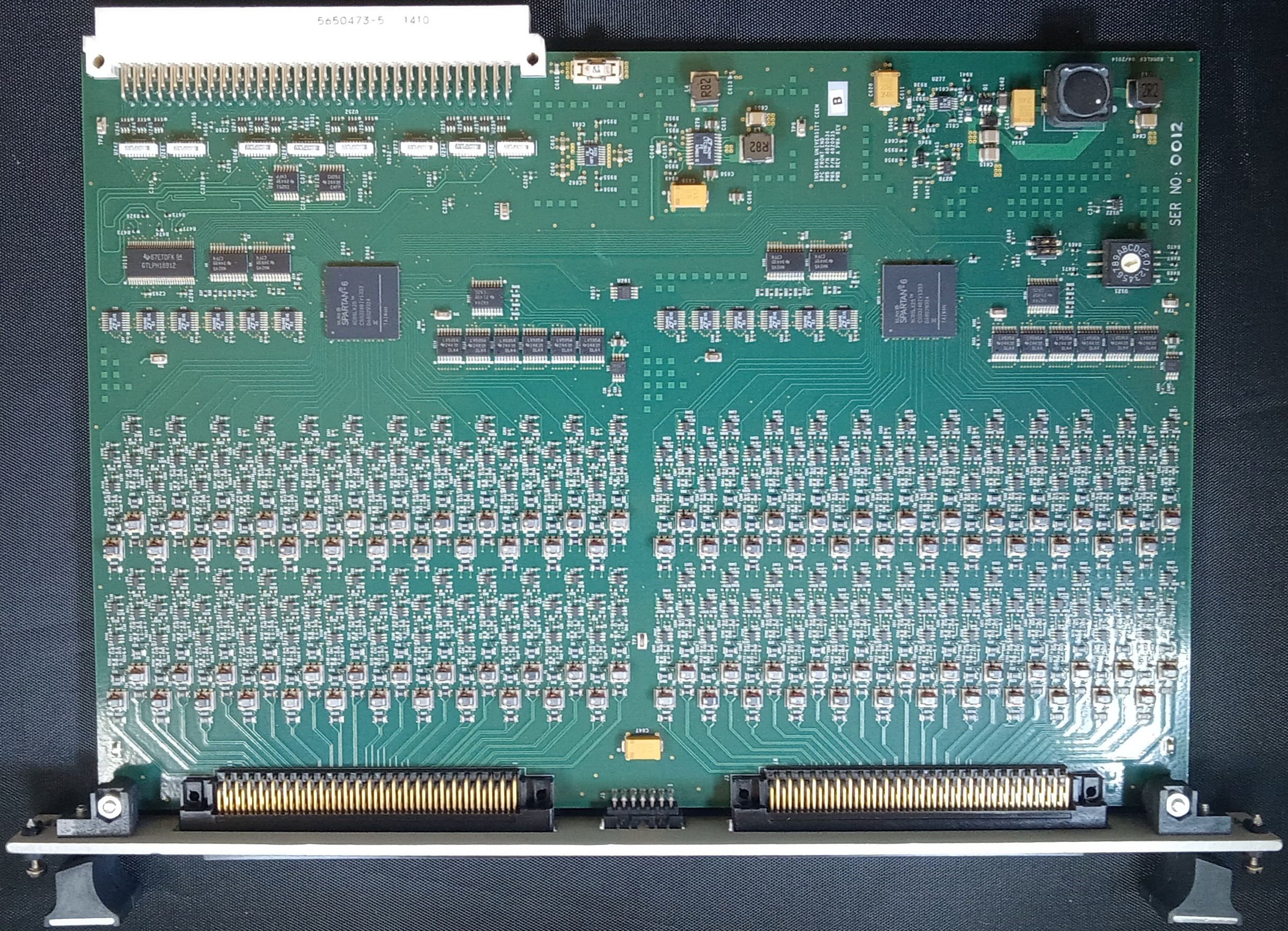}  
  \includegraphics[angle=90, width=0.49\columnwidth]{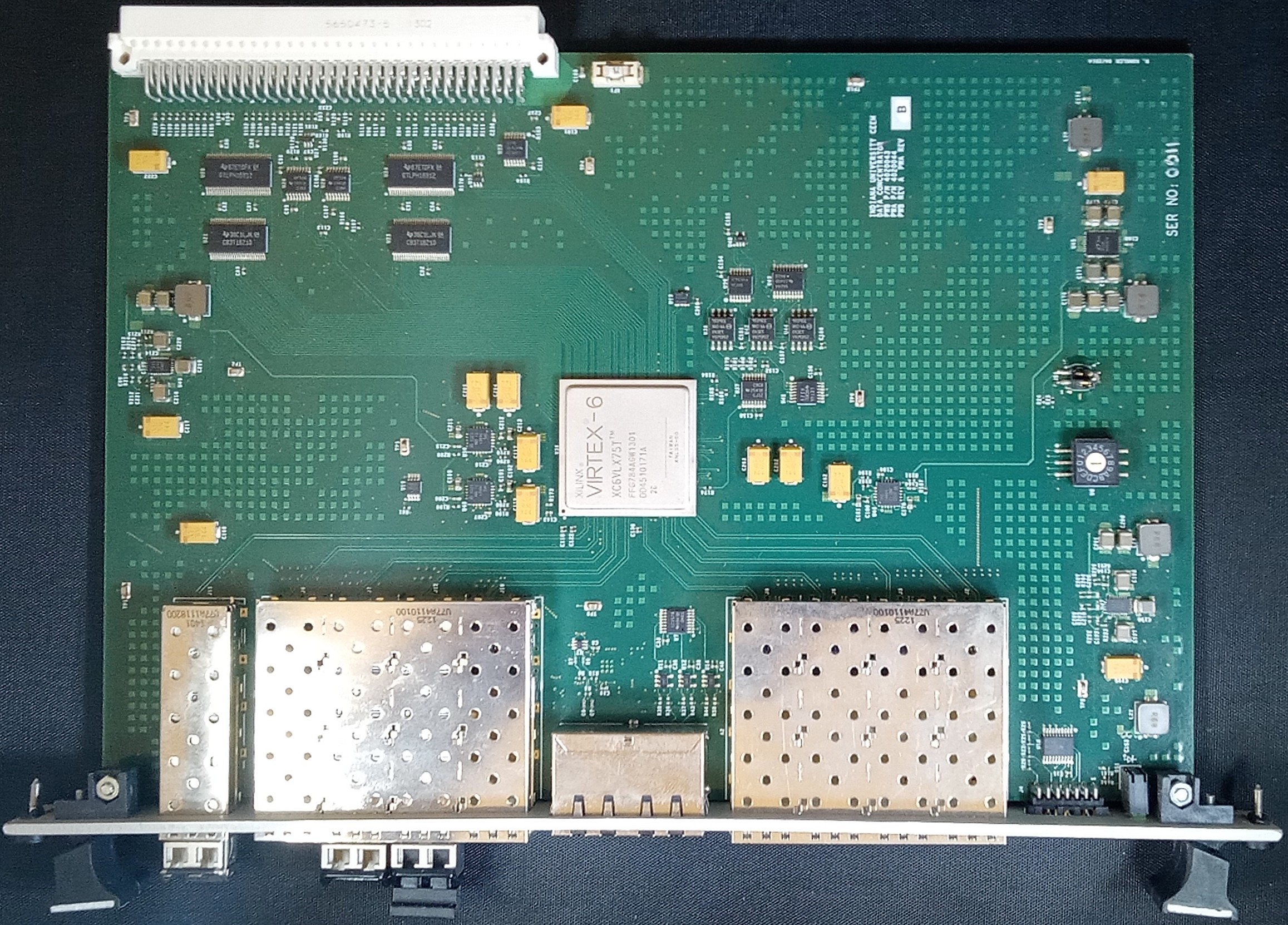}
  \caption{Left: RPC front-end board. Right: Data Concentrator.}
  \label{fig:RPC_FEB_and_DataCon}
\end{figure}
\begin{figure*}
  \centering
  \includegraphics[width=5.4in]{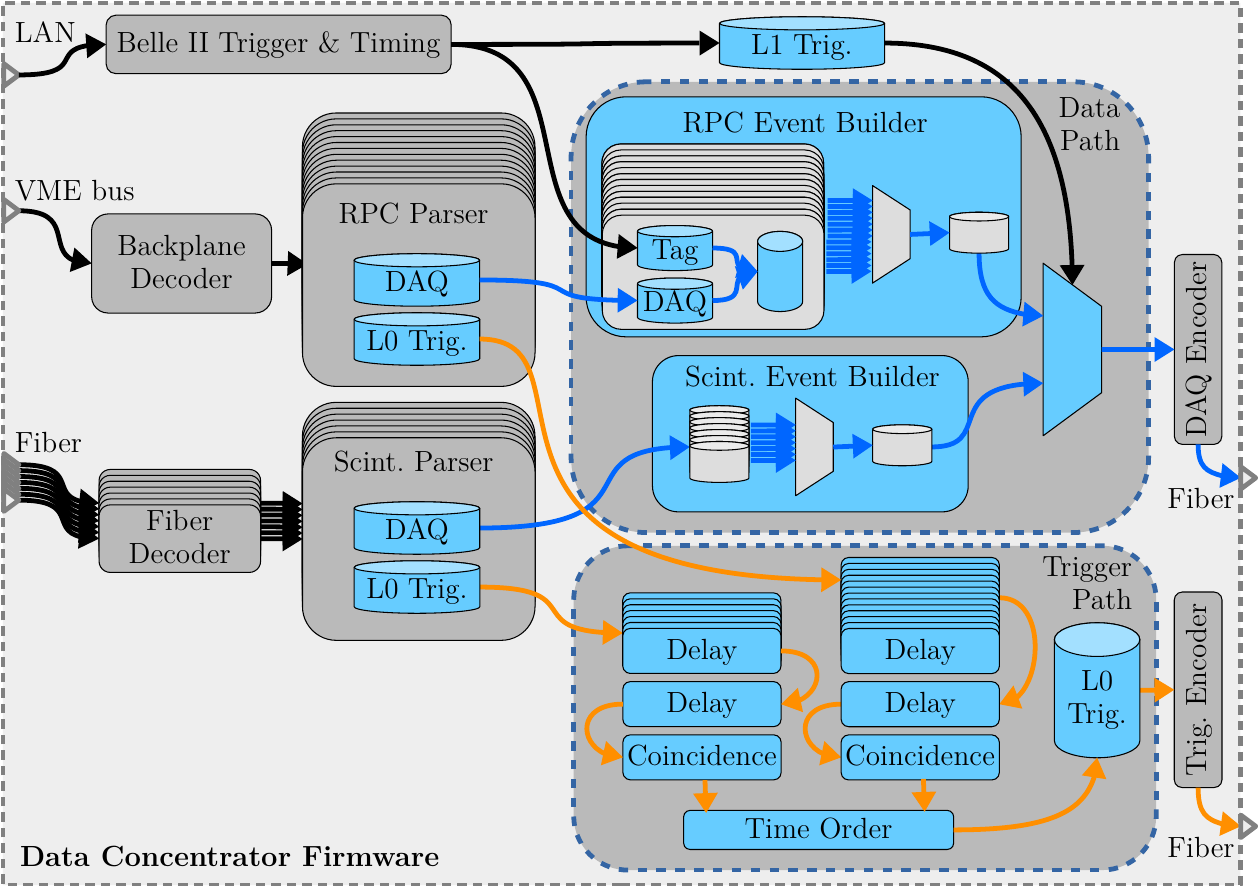}
  \caption{Flow chart depicting the top level of the Data Concentrator firmware.
           The blue (orange) arrows represent the data path (trigger path).
           Clock distribution, register interface, and fiber transceiver interface omitted.}
  \label{fig:DataCon_fw}
\end{figure*}

Each Data Concentrator (Fig.~\ref{fig:RPC_FEB_and_DataCon}) contains nine serial fiber transceivers, two RJ45 connectors,
and a single FPGA.\footnote{Xilinx Virtex-6 XC6VLX75T}
It collects data packets from either two scintillator modules over fiber and thirteen RPC modules over a VME backplane in the barrel region, or from six to seven scintillator modules (all over fiber) in the endcap regions.
Two additional fiber transceivers are used to transmit hit packets to the Belle~II 
data acquisition system and to transmit trigger packets to the Belle~II trigger 
decision system, respectively. 
One RJ45 provides clock and trigger inputs while the other is used for FPGA programming.
A block diagram of the Data Concentrator firmware is depicted
in Fig.~\ref{fig:DataCon_fw}.

For any particular L1 trigger, the Data Concentrator gathers any RPC packets it has and then
waits for each connected SCROD to respond with either a valid or null data packet. The combined data packets
are then sent to the Belle~II data acquisition (DAQ) system via fiber using a custom protocol called Belle2Link.

For sending configuration data packets to the front-end electronics, the Belle2Link protocol is also used. Further,
to send data to the SCRODs, the Data Concentrator translates everything from 
Belle2Link and passes it on to the SCROD via the Xilinx Aurora protocol.

\section{Calibration of the TARGETX ASICs and SiPMs}
The TARGETX ASIC has 61 registers for device calibration. Fifteen registers set
the width of the trigger bit pulses from the one-shot circuits, another
15 set the trigger threshold comparators for each channel, and one register enables a test pattern.
The remaining
30 registers are used to optimize the performance of the TARGETX.  
Of these 30 remaining registers, some are for tuning the shape of the Wilkinson
ramp, some for timing of sample storage and addressing, and some for time-base
corrections. The initial tuning of these registers is described in Ref.~\cite{Edralin2016}.
Further, on the RHIC, there is one \SI{5}{\V} 8-bit DAC for fine-tuning the SiPM gain on each channel.

\subsection{Wilkinson Ramp Tuning}\label{section:ramp_tuning}
The TARGETX ASIC uses a Wilkinson ADC to digitize its analog samples. 
During digitization, one input of a comparator is driven by the sample
cell that is being digitized and the other by a linearly increasing voltage source (the Wilkinson ramp).
The time it takes for the ramp voltage to reach the sample voltage is proportional
to the sample voltage.
When the ramp begins, a 12-bit Gray-code counter is activated.
When the ramp exceeds the sample voltage, the comparator output latches the Gray-code counter
to its present value. We refer to this value as the number of Wilkinson ADC counts ---
a digital value that corresponds to the voltage of the sample. 
The Wilkinson clock that increments the Gray-code counter is provided over low-voltage differential signals
generated by the SCROD FPGA.

\begin{figure*}
  \centering
  \includegraphics[width=0.47\linewidth]{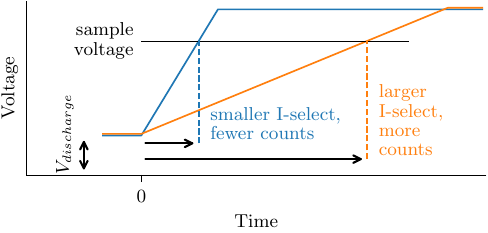}\hspace{0.05\linewidth}
  \includegraphics[width=0.47\linewidth]{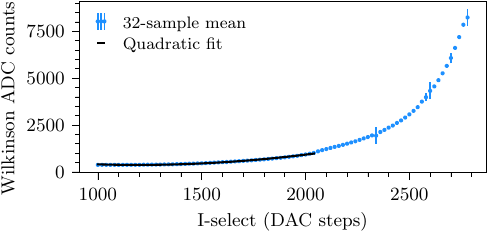}
  \caption{Left: Wilkinson ADC principle. A counter is enabled when the ramp
           begins and continues counting until the ramp
           exceeds the sample voltage. The discharge voltage of the ramp
           generator sets the starting voltage. Lower values of I-select make steeper
           slopes, which leads to shorter digitization times and fewer
           Wilkinson ADC counts.
           Right: Measured baseline as a function of ramp slope. 
           Baseline refers to a constant-voltage input without signal present. Quadratic behavior  
           reflects saturation mode of a p-type FET at drive strengths below $V_D/2$.
           The 12-bit Gray-code counter overflows twice near the end of the scan.
           The data was overflow corrected at intervals of $2^{12}$, hence the larger error bars near
           these boundaries. Also, outlier samples are more common near $2^{11}$ ADC counts hence
           the remaining larger error bars.}
  \label{fig:ramp_tuning}
\end{figure*}

In the TARGETX ASIC, the I-select register controls the slope of the Wilkinson ramp,
and the V-discharge register controls the starting voltage of the ramp (Fig.~\ref{fig:ramp_tuning}--left).
Increasing I-select decreases the slope. We measure the behavior of I-select
by turning off the HV and digitizing samples corresponding to the input offset voltage
of the TARGETX channels, which is set at about 3/4 of the dynamic range for TARGETX inputs. We 
measure the mean number of Wilkinson ADC counts for groups of 32 samples at 
every I-select setting. At higher values of I-select, and with a Wilkinson
clock period of \SI{7.87}{ns}, the input offset voltage
is high enough that the Wilkinson counter exceeds its maximum and continues
counting from zero. In the measurement, we correct this overflow in software  (Fig.~\ref{fig:ramp_tuning}--right).
The quadratic
shape below V$_D$/2 is characteristic of the P-channel MOSFET that is driving the 
ramp generator. We select a value for I-select within this quadratic band
that maximizes the dynamic range of the Wilkinson ADC without overrunning it.

One unfortunate design flaw of the preamplifiers is that they were optimized
for single-pixel detection, yet they saturate quite easily when a large number
of SiPM pixels fire in tandem, causing the negative-going SiPM pulse to have a flat
bottom rather than a well-defined extremum. This hardware choice could not be changed,
so with this in mind, we set V-discharge so that saturated pulses 
will have a minimum near zero Wilkinson ADC counts. This choice allows
for quicker digitization times.

\subsection{Aligning the Trigger Threshold Baseline}
The analog input of each TARGETX channel is not only sampled by the sampling array
but also provides one input of the trigger-threshold comparator. 
The other input is provided by a 12-bit DAC. These trigger-threshold DAC settings
are each tuned independently.
Nominally, all of the 15 analog inputs would have the same DC offset, and the trigger-threshold DACs 
would be identical.
In practice, however, there is variance. 

\begin{figure}
  \includegraphics[width=0.9\columnwidth]{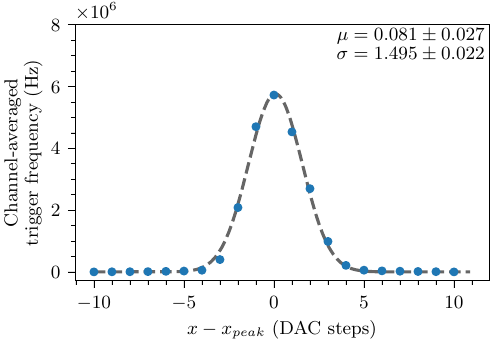}\\
  \includegraphics[width=\columnwidth]{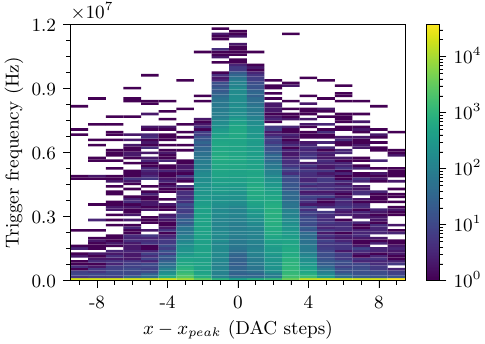}
  \caption{The variable $x$ refers to the trigger-threshold DAC setting of any particular channel.
           Top: sum of all baseline scans divided by number of installed channels. 
           Bottom: histogram of all the scans in order to demonstrate the variance
           between all installed TARGETX channels.}
  \label{fig:baseline_scan}
\end{figure}

The first step to aligning all channels is to turn the HV off and measure the frequency
of trigger bits versus the trigger-threshold DAC setting for each channel (Fig.~\ref{fig:baseline_scan}). This is done in firmware
by counting the number of trigger bits within a programmable time interval, then reading
out the count via the status register interface. The result is a normal distribution
centered on the value of interest.
The average width of all the distributions is \num{1.495 \pm 0.022} trigger DAC steps.

This measurement provides the trigger threshold that corresponds to zero pixels hit on the SiPM.
We call this the trigger-threshold baseline value.
This measurement is performed on all 18,560 installed channels and saved to a database.
A histogram of the measured baseline values for all channels is shown on the left of Fig.~\ref{fig:scan_summaries}. 
Ultimately, we want to tune the trigger threshold to a value that corresponds to a predetermined
number of fired pixels from the SiPM.
However, the voltage seen by a single pixel depends on the gain of the SiPM.

\subsection{Coarse Gain Adjustment}
Before finely tuning the gain on more than 18,000 channels, it is prudent to first perform a coarse
measurement to get close to the desired value. According to the SiPM vendor, at a \SI{70}{\V} bias,
the frequency of SiPM pulses larger than 1.2 pixels is \SI{75}{\kHz}, and \SI{750}{\kHz} for 0.5 pixels,
respectively. 
Extensive testing on a few channels pinned down a trigger-threshold DAC setting of 35 below the baseline value
as corresponding to a trigger threshold of 1.2 pixels. 

\begin{figure}
  \centering
  \includegraphics[width=\columnwidth]{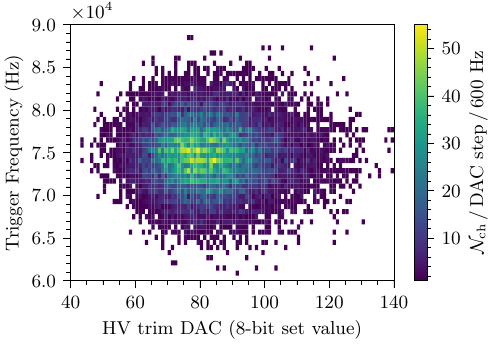}
  \caption{
           Histogram of coarse HV-trim DAC adjustment and corresponding
           trigger scalar frequency for each (when the trigger thresholds are set to 35 trigger DAC steps below the baseline value).}
  \label{fig:scan_summaries}
\end{figure}

Because the low side of the SiPM bias voltage is set with an 8-bit \SI{5}{\V} DAC, we set
the high side of all SiPMs to \SI{73}{\V} --- about \SIrange{1}{5}{\V} above the breakdown.
Now, with knowledge of the trigger-threshold baseline value from the previous section,
we set each channel's trigger threshold DAC to 35 below its baseline value, and we measure
the trigger bit frequency versus the HV-trim DAC setting. We select the DAC setting
that yields a trigger-bit frequency closest to \SI{75}{\kHz} as the coarse setting.
A histogram of the best HV-trim DAC values and corresponding frequencies for all channels is shown on the right of Fig.~\ref{fig:scan_summaries}.

\subsection{Normalizing Gain on All SiPMs}
\begin{figure}
  \begin{center}
    \includegraphics[width=\columnwidth]{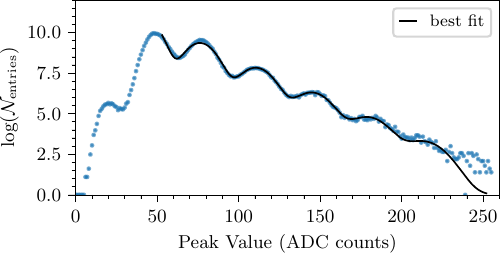}
    \caption{Example of darkrate-based single-photon spectrum. The trigger threshold is set to about
    one PE, which has the effect of diminishing the first two photo peaks.
    The best fit to a function with six peaks is drawn on top of the data.}
    \label{fig:one_sps}
  \end{center}
\end{figure}

Single-photon spectra (SPS) are one way to measure gain (Fig.~\ref{fig:one_sps}). 
We use a histogram of pulse height measurements for this. 
Each peak in the spectrum corresponds to a number of pixels fired.
The separation between adjacent peaks is the gain of the SiPM plus preamplifier.
To set the gain uniformly on all channels, we need to know the gain as a function of the 
HV-trim DAC setting. This requires many SPS and many fits. 
Ordinarily, this is achieved by using a calibration source such as an LED, and a climate
chamber can allow for testing at different temperatures.

The KLM detector was designed without calibration sources in mind. This leaves the inherent dark rate 
of the SiPM as the only means of measuring gain. A further complication is encountered
in the data acquisition system: the Data Concentrator firmware and the readout PC were
designed to accept 8-byte packets from the scintillator front end, not waveforms.
Finally, whatever procedure is used, it must be repeated on all 18,560 channels.
To overcome these issues, we record SPS in firmware by keeping a histogram of waveform peak measurements
in the FPGA's block RAM. Each address in the allocated RAM corresponds to one bin of the histogram, and
filling the histogram just requires reading the RAM, adding 1, and writing again.

To make the measurement, a special reset signal clears the SPS RAM contents, 
and trigger thresholds are turned off for all but one channel on the motherboard, which is
set at a threshold of about one photoelectron. The firmware is configured in self-triggering
mode, and pulse-height data is collected for 90 seconds. In this way, all 150 channels on a motherboard
can be measured in 225 minutes. Offline software executes the measurement on all
installed motherboards in parallel. 

While the experiment hall is climate-controlled, 
some daily and seasonal variation in temperature is expected. Seasonal variation
is mitigated by performing the calibration all at once. The absolute gain will fluctuate
due to seasonal variation, but normalization across all channels should not be affected.
Daily variation in temperature cannot be mitigated. 

After each \SI{90}{s} measurement, the RAM contents are read out using the register interface for offline analysis. 
The procedure is repeated at different bias voltages over the range of \SIrange{2}{3}{\V} past breakdown
to establish gain as a function of bias voltage for each channel. 

We fit each SPS using a function that is a sum of Gaussian distributions which are regularly spaced by the parameter $a_0$ and whose amplitudes 
decrease by a power law $\chi^k$, $0 < \chi < 1$:
$$f(x) = A\sum_{k=1}^{N_{PE}}\chi^k e^{\frac{-(x - (x_0 + k a_0))^2}{2(\sigma_0 + k\sigma_1)^2}}\;,$$
where $\chi$ is the optical crosstalk probability,
$x_0$ is a pedestal offset,
$\sigma_0$ describes electronic noise and ADC resolution,
and $\sigma_1$ scales with the number of pixels fired.
The only purpose of the fit is to extract the parameter $a_0$, i.e., the gain in units of ADC counts per photoelectron.
The power law stems from crosstalk between pixels. This is due to infrared photons created in the
avalanche. If these photons are emitted isotropically, then the probability that they will
hit any other pixels is given by a power law. The crosstalk probability also depends on how many infrared photons
are created in a typical avalanche, and on the quantum efficiency for converting infrared
photons to photoelectrons. These can all be lumped into the parameter $\chi$ in the fit.

\begin{figure}
  \centering
  \includegraphics[width=\columnwidth]{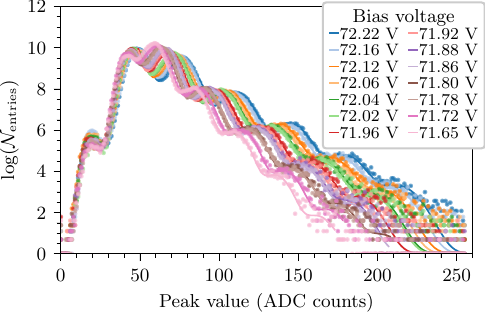}\\
  \includegraphics[width=\columnwidth]{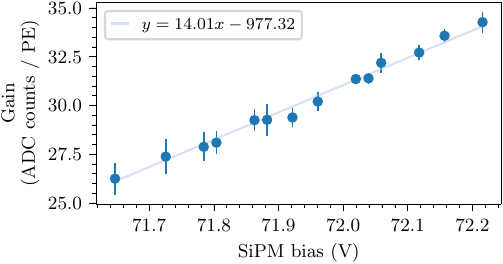}
  \caption{Top: Single-channel example of multiple gain fits at different SiPM bias voltages. Bottom: Mean separation between peaks, $a_0$, as a
  function of bias voltage.}
  \label{fig:one_gain_fun}
\end{figure}

To measure gain as a function of bias voltage,
we repeat the above procedure for different bias voltages (Fig.~\ref{fig:one_gain_fun}).
For 10 data points per channel, we have to perform 185,600 fits. A non-linear least-squares minimizer is used. The fitting procedure is optimized by
studying many examples. The best results are achieved by fitting the logarithm of the number of entries to log$(f(x))$, 
and by providing the minimizer with the loss function $\rho(\delta^2) = 2(\sqrt{1 + \delta^2} - 1)$, where $\rho$ is the loss and $\delta$ is a residual.
These choices improve the sensitivity of the fit to $a_0$, the only parameter of interest. 

\begin{figure}
  \centering
  \includegraphics[width=0.925\columnwidth]{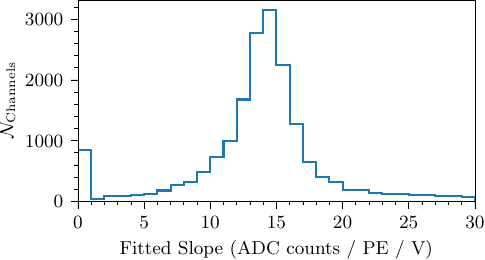}\\
  \includegraphics[width=0.925\columnwidth]{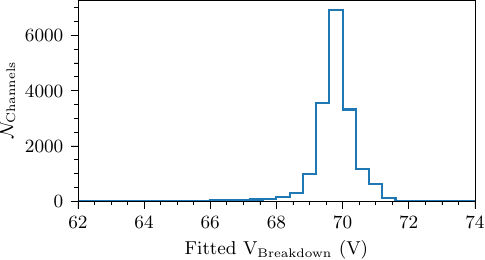}
  \caption{Results of SiPM gain calibration. Top: Slope of ADC counts\,/\,photoelectron\,/\,V
           for all installed SiPMs. The peak at zero corresponds to channels for which 
           converging fits are not obtained using the automated measuring and fitting setup.
           For these channels, the coarse
           HV-trim DAC setting is retained. Bottom: $x$-intercept of each linear fit. This
           is the measured breakdown voltage of each SiPM}
  \label{fig:gain_summary}
\end{figure}

The result of this procedure on all 18,560 channels is a converging linear fit on all but approximately 1,000 channels (Fig.~\ref{fig:gain_summary}).
The mean gain slope is 15 ADC counts\,/\,PE\,/\,V, and the mean breakdown voltage ($x$-intercept) is around \SI{70}{\V}.
After the fit procedure, for all channels with a converged linear fit, the required HV-trim DAC value
needed to achieve 30 ADC counts\,/\,PE is calculated using the linear fit function, and these
values are written to the KLM detector's configuration database. For the channels without a converged fit,
the HV-trim DAC value from the coarse gain adjustment is retained. 

\section{Performance}
\begin{figure*}
  \centering
  \includegraphics[width=\linewidth]{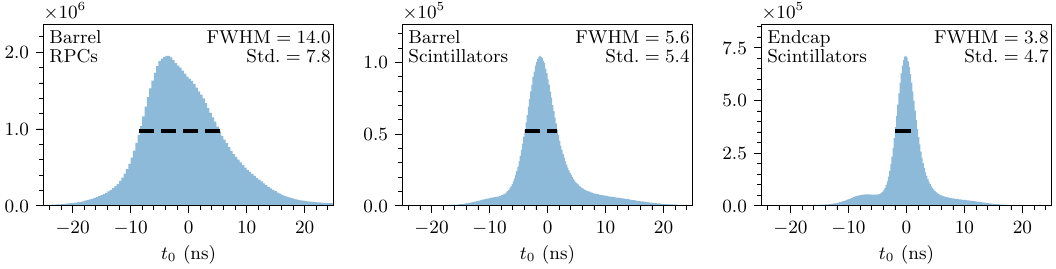}
  \caption{KLM hit time distributions for RPCs (left), barrel scintillators (middle), and endcap scintillators (right). The full width at half maximum (FWHM) and 
  the standard deviation are indicated in each plot.}
  \label{fig:time_calibration}
\end{figure*}
The hit time recorded by the firmware is delayed compared to the actual time that a particle interacts with a KLM module.
This is due to the signal transmission from the hit position to the readout electronics.
Because of the high background level expected at Belle~II's design luminosity, precise measurement of hit time ($t_{0}$) resolution is necessary.
Precise timing will lead to better reconstruction of tracks and $K_{L}$ clusters.
For a single-channel hit in the KLM detector, we define $t_{0}$ as,
\begin{center}
$t_{0} = T_{\rm{rec}} - (T_{0} + T_{\rm{flight}} + T_{\rm{propagation}} + T_{\rm{collect}} + T_{\rm{cable}})$.
\end{center}
Here, $T_{\rm{rec}}$ is the time of the hit recorded by the firmware and is important for reconstructing 2D hits and subsequently tracks.
The $e^+e^-$ collision time for each event is $T_{0}$ (EventT0).
We use EventT0 information from the Central Drift Chamber (CDC).
The variable $T_{\rm{flight}}$ denotes the flight time of the particle from the interaction point to the hit position in the KLM module.
For each recorded hit, $T_{\rm{flight}}$ is obtained by matching the KLM detector hit position to the extrapolated hit from CDC, which indicates the relative distance from the beam line.
The variable $T_{\rm{propagation}}$ is the charge propagation (photon propagation) time for RPCs (scintillators) between the hit position and the end of the strip (fiber).
In the scintillators, it is estimated using $L_{\rm{propagation}}/c_{\rm{eff}}$, where $L_{\rm{propagation}}$ is the propagation distance of the signal and $c_{\rm{eff}}$ is the effective speed of light in the fiber.
The time between photoelectric conversion in a SiPM and leading-edge timestamping is $T_{\rm{collect}}$ and depends on the number of pixels fired in the SiPM.
This contribution is small and currently treated as a constant.
Finally, $T_{\rm{cable}}$ is the transmission time of the signal over the ribbon cables.
The time delay of KLM detector hits is mainly from the ribbon cables, which is different for each strip due to the detector structure.

The distribution of $$T_{\rm{cable}} = T_{\rm{rec}} - (T_{0} + T_{\rm{flight}} + T_{\rm{propagation}})$$ for each strip is fitted with a Gaussian function.
The mean value of these Gaussian distributions is fitted with a constant function to obtain the weighted global mean.
The difference between the mean value of one strip and the global mean value is used as the calibration constant for that strip and is stored in the Belle~II conditions database\cite{Ritter-condDB}.
In the Belle~II software framework\,\cite{Kuhr_2018, the_belle_ii_collaboration_2022_6949513}, $T_{\rm{cable}}$ values are subtracted when reconstructing the hits.

The results of the time calibration for data collected at the beginning of 2024 are shown
in Fig.~\ref{fig:time_calibration}. For $t_{0}$ resolution, we state the standard deviation and full width at half maximum (FWHM).
The $t_{0}$ resolutions (FWHMs) for RPCs, barrel scintillators, and endcap scintillators are \SI{7.8}{ns} (\SI{14.0}{ns}), \SI{5.4}{ns} (\SI{5.6}{ns}), and \SI{4.7}{ns} (\SI{3.8}{ns}), respectively. The tails and asymmetric shapes of these distributions are likely due to a combination of the calibration
algorithm itself and background hits. Currently, the algorithm does not account for the
curvature of charged tracks within the Belle~II magnetic field, for scintillator hits in which waveform digitization was skipped, or for scintillator hits in which the pulse never exceeded the threshold for leading-edge time determination.

\section{Conclusion}
An electronic readout system, combining data acquisition for two distinctly different detector technologies,
was designed, installed, and commissioned for the Belle~II KLM subsystem. A challenging
task was the creation of readout firmware for the scintillator readout that utilizes
the waveform-digitization feature of the TARGETX ASIC at trigger rates
up to \SI{30}{\kHz}.
Having achieved this goal, we anticipate an improvement in particle identification for the KLM subsystem.
With the waveform readout now working, we can disambiguate multi-channel hits from a single ASIC, leading to improved track and cluster resolution. 
Waveform feature extraction has the potential to improve the hit-time resolution
to about \SI{1}{\ns}, which will help with background rejection. Knowledge of SiPM
pulse heights may also unlock new analysis techniques, perhaps incorporating pulse-height
measurements to resolve
low momentum muons from punch-through pions which made it into the KLM detector.

A second challenge---to calibrate gains on more than 18,000 SiPMs that
were installed without any calibration sources---was resolved successfully.
We deployed a procedure for efficiently recording
single-photon spectra in firmware, and we developed an automated fitting procedure
to extract the parameter of interest (the gain). We used the calibration results
to homogenize the gain on more than 17,000 of the installed SiPMs. 

Future work is required to improve the stability of the waveform digitization. 
Improvements in the fitting procedure or recollecting single-photon spectra on
about 1,000 channels whose fits did not converge should be considered. 
Additionally, a more sophisticated feature-extraction algorithm, such as
a finite impulse response filter,  may lead
to further improvements in time and peak resolution. 
\section*{Acknowledgments}
We would especially like to thank Brandon Kunkler and Gary Varner
who were instrumental to the success of this project from the very beginning.

This work, regarding the Belle II detector, which was built and commissioned prior to March 2019,
was supported by
%
the National Key R\&D Program of China under Contract No.~2022YFA1601903 and
the National Natural Science Foundation of China and Research Grant
No.~12175041;
%
the Istituto Nazionale di Fisica Nucleare and the Research Grants BELLE2;
%
the HSE University Basic Research Program, Moscow;
%
the U.S. National Science Foundation and Research Grant
No.~PHY-1913789 
and
and the U.S. Department of Energy and Research Awards
No.~DE-AC06-76RLO1830, 
No.~DE-SC0009973, 
No.~DE-SC0010007, 
No.~DE-SC0010504, 
No.~DE-SC0012704, 
No.~DE-SC0019230, 
No.~DE-SC0021430, 
and 
No.~DE-SC0022350. 
These acknowledgements are not to be interpreted as an endorsement of any statement made
by any of our institutes, funding agencies, governments, or their representatives.

We thank the SuperKEKB team for delivering high-luminosity collisions;
the KEK cryogenics group for the efficient operation of the detector solenoid magnet and IBBelle on site;
the KEK Computer Research Center for on-site computing support; the NII for SINET6 network support;
and the raw-data centers hosted by BNL, DESY, GridKa, IN2P3, INFN, and the University of Victoria.

\bibliographystyle{elsarticle-num} 
\bibliography{reference_list.bib}

\end{document}